# On the Space-Time Statistics of Motion Pictures

**Dae Yeol Lee,[1] Hyunsuk Ko,[2] Jongho Kim,[3] and Alan C. Bovik[1,*]**

[1]*Department of Electrical and Computer Engineering, The University of Texas at Austin, Austin, TX, USA*
[2]*Division of Electrical Engineering, Hanyang University ERICA, Ansan, Republic of Korea*
[3]*Media Coding Research Section, Electronics and Telecommunications Research Institute, Daejeon, Republic of Korea*
*\*bovik@ece.utexas.edu*

**Abstract:** It is well-known that natural images possess statistical regularities that can be captured by bandpass decomposition and divisive normalization processes that approximate early neural processing in the human visual system. We expand on these studies and present new findings on the properties of space-time natural statistics that are inherent in motion pictures. Our model relies on the concept of temporal bandpass (e.g. lag) filtering in LGN and area V1, which is similar to smoothed frame differencing of video frames. Specifically, we model the statistics of the differences between adjacent or neighboring video frames that have been slightly spatially displaced relative to one another. We find that when these space-time differences are further subjected to locally pooled divisive normalization, statistical regularities (or lack thereof) arise that depend on the local motion trajectory. We find that bandpass and divisively normalized frame-differences that are displaced along the motion direction exhibit stronger statistical regularities than for other displacements. Conversely, the direction-dependent regularities of displaced frame differences can be used to estimate the image motion (optical flow) by finding the space-time displacement paths that best preserve statistical regularity.

## 1. Introduction

Among the foundational concepts underlying modern visual neuroscience are the ideas that our visual neurons have evolved to match the statistical properties of the visual environment, and that the early visual system exploits these properties to remove redundancies that are present in sensory inputs [1]. Many studies have investigated the statistical properties of spatial visual signals and how they have formed, and are processed by, front-end processing in vision systems [2,3]. Other studies have described statistical regularities that are present in natural images that have been subjected to bandpass decomposition followed by local contrast normalization and have related them to the response properties in retinocortical neurons [4-10]. Quantitative models of natural picture statistics have been used with great success in the creation of perceptual algorithms for image processing, such as picture denoising [11] and picture quality prediction [12-18].

However, to date, less work has been done on modeling the temporal statistics of motion pictures. While some work has been done on quantifying the statistical properties of videos [19,20], this work has been limited to the study of simple frame difference, but without exploring space-time statistical properties that may arise from motion. Towards advancing progress in this direction, we develop new models of motion picture statistics in space-time. Specifically, we model differences between motion picture frames that are relatively displaced in both space and time, and are then divisively normalized by locally pooled energies of the displaced differences. We make a number of interesting observations, one of which is that space-time displaced and normalized differences are more statistically regular along motion trajectories. We make use of this by showing that good motion predictions (optical flow vectors) can be obtained by finding space-time directions of maximum regularity.

The remainder of the paper is organized as follows. We revisit well-established statistical models of natural pictures, and in Section 2, motivate expanding these theoretical models into space-time. Section 3 describes the way we model space-time statistics, by first constructing frame-difference volumes defined by space-time displacement trajectories, followed by volumetric divisive normalization in the space, time, and space-time directions. We empirically reveal certain statistical regularities of the space-time frame-differences displaced along local motion trajectories. It is plausible that the vision system exploits these regularities to help sense local motion flow. In Section 4, we explore this proposition, by developing a simple motion prediction model based on a measure of space-time directional regularity. Surprisingly, the accuracy of the model is comparable to that of popular and canonical optical flow algorithms. We discuss our findings and their possible ramifications for visual motion sensing, as well as applications, in Section 5.

## 2. Background

A variety of studies have explored the deep relationships between front-end processing in the vision system and the statistical laws that govern natural images. Natural picture statistics are believed to have deeply impacted the way that vision systems have evolved to transform and encode visual stimuli, while extracting information relevant to scene understanding.

### 2.1 Statistical regularity of natural pictures

The most widely accepted models of natural pictures statistics involve linear bandpass decompositions, followed by a non-linear divisive normalization.

Bandpass decomposition models are generally motivated by scale and/or orientation-sensitive decorrelating processes occurring in retina and/or area V1. As discussed in Field et al. [21], multi-channel bandpass picture decompositions that are learned on large sets of natural images reveal sparse, highly efficient representations, which are similar to the responses of cortical receptive fields [8,22]. Some of the most widely used decomposition models are the steerable pyramid [23] and local mean subtraction [4]. Fig. 1-(b) depicts the empirical distribution (histogram) of the coefficients obtained from local mean subtraction. The histogram has a peaky, heavy-tailed, approximately symmetric shape, which is similar to a generalized Gaussian distribution (GGD).

Band-pass filtered image signals subjected to divisive normalization mirror the nonlinear behavior of retino-cortical neurons. For example, nonlinearities of V1 neurons control the gains of neuronal responses by scaling them by the (weighted) average energy responses of neighboring neurons over frequency, orientation, space, and time [5,6,9,10]. Divisive normalization further regularizes the bandpass responses by further decorrelating and Gaussianizing them.

More than 25 years ago, Ruderman conducted an experiment on a collection of naturalistic (photographic) pictures, whereby he differenced pixel luminances with local averages of neighboring luminances, then divided them by local RMS contrasts [4]. The resulting images exhibited a remarkable degree of statistical regularity that has been borne out by numerous studies. Specifically, pictures that are processed in this way exhibit strongly gaussianized histograms, as illustrated in Fig. 1-(c), and are substantially spatially decorrelated. For brevity, we shall refer to the resulting processed values as mean subtracted, contrast normalized (MSCN) coefficients. MSCN coefficients of the various types of frame differences will form a basic tool of our analysis. Other decompositions, such as steerable pyramids followed by divisive normalization could also be used, but MSCN coefficients are simple, effective, and computationally efficient. The conventional form of MSCN coefficients computed on an image is given by

$$\hat{I}(i,j) = \frac{I(i,j) - \mu(i,j)}{\sigma(i,j) + C}, \qquad (1)$$

where $I$ refers to an input image, $i \in [1, w]$, and $j \in [1, h]$, where $w$, and $h$, are the input image dimensions, $C$ is a stabilizing (saturation) constant, and $\mu$ and $\sigma$ are locally gaussian-weighted mean and contrast computed as

$$\mu(i,j) = \sum_{l=-L}^{L} \sum_{m=-M}^{M} \omega_{lm} I(i+l, j+m) \qquad (2)$$

and

$$\sigma(i,j) = \sqrt{\sum_{l=-L}^{L} \sum_{m=-M}^{M} \omega_{lm} [I(i+l, j+m) - \mu(i,j)]^2}, \qquad (3)$$

respectively. The window function $\omega_{lm}$ is a 2D symmetric Gaussians sampled out to three standard deviations ($L = 5$, $M = 5$) and rescaled to unit volume. The numerator of (1) serves as a bandpass operation that shapes the distribution of the transformed coefficients to GGD, as shown in Fig. 1-(b), and the denominator of (1) serve as a divisive normalization operation that gaussianizes the distribution of coefficients, as shown in Fig. 1-(c).

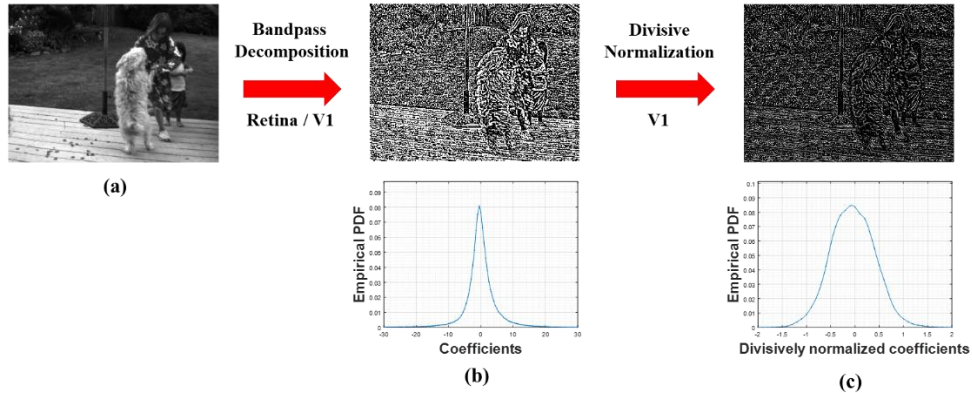

Fig. 1. Statistics of natural pictures: (a) natural image; (b) plot of distribution of band-pass coefficients before divisive normalization; (c) plot of distribution of the same band-pass coefficients after divisive normalization.

*2.2 Motivating space-time displaced frame-differences*

Motion pictures (or videos) may be regarded as sequences of still pictures that sequentially project onto the retinae. However, simply aggregating visual information drawn from sequences of still images is inadequate to explain temporal aspects of visual perception. While existing statistical models of videos have been developed for simple frame differences [19,20], these do not account for space-time directional statistics related to motion.

While it is true that models of temporal bandpass processing from retina through lateral geniculate nucleus (LGN) and primary visual cortex accomplish significant reductions of spatial and temporal redundancies [24], the significant literature on spatial picture statistics and the less developed body of knowledge on temporal motion picture statistics do not easily extend to space-time, because of the complexities of object motion [25]. Our approach to the problem is different, as, instead of attempting to find regularities of motion vectors, or in purely temporal or spatial (or product combinations thereof) bandpass filter responses, we instead study the statistics of space-time displaced frame-differences. This can be motivated by the fact that most projected object motions or head/camera motions are small over short time periods, hence space-time displacements of frames would be expected to reveal redundancies in the motion direction. However, this topic may also relate to very small eye movements (microsaccades), which are theoretically implicated in more efficient visual encoding processes [26,27]. These encoding efficiencies would likely involve exploiting redundancies in the visual signal over small space-time displacements. One simple way to probe space-time statistics and to model redundancy-reduced visual signals is by computing and analyzing normalized bandpass differences over varying space-time displacements.

## 3. Space-time motion picture statistics

In the following, we describe our statistical modeling methodology which we use to probe potential statistical regularities of visual data in space-time. First, we describe the framework we use to compute displaced space-time frame differences, followed by the way we model and compute divisive normalization.

### 3.1 Space-time displaced frame-differences

Here we explain the procedure we used to construct displaced space-time frame-differences, which are defined in regards to a set of space-time displacement trajectories. These are intended to capture local dynamic space-time processes including those that are static in time, or primarily temporal (such as local flicker), or that reflect small space-time object and/or eye/head/camera movements. These displacements may also arise from small, microsaccadic eye movements that may be implicated in improving encoding efficiency in the early vision system.

Consider the luminance values of video frames, which we denote as $\{I_t; t=1,\ldots,T\}$, where $T$ is the total number of frames in a given clip. Then, given a 3D space-time displacement vector $d = (x, y, t)$, define the displaced difference between frames $t_0 + t$ and $t_0$ at each $(i, j)$ by

$$FD(i, j)_{(d,t_0)} = I_{t_0}(i, j) - I_{t_0+t}(i+x, j+y). \qquad (4)$$

Note that the indices $(x, y, t)$ are constrained by the finite extent of the video clip according to $i \in [\max(1, 1-x), \min(w, w-x)]$, $j \in [\max(1, 1-y), \min(h, h-y)]$, and $t \in [1, T-t_0]$, where $h$ and $w$ refer to the spatial height and width of the video, respectively.

For clarity, we will refer to frame-differences between temporally adjacent frames for which the relative spatial displacements between the two involved frames are zero, as "non-displaced." While we will allow for different degrees of temporal displacements, we will still refer to differences between spatially aligned frames separated by single time units (e.g. 1/30 or 1/60 sec) in this way. While the statistics of time-displaced frames have been studied before, here we extend the analysis by studying the statistical properties of more general spatially and temporally displaced frame differences that occur along space-time trajectories. We consider the following three types of space-time patch trajectories, with one of our aims being to understand whether the differences in statistical properties that are associated with, and that potentially distinguish the trajectory types:

- **Motion:** Space-time trajectories of patches in the direction of (ground-truth) motion

- **Non-displaced:** Static, spatial patches without motion

- **Random:** Patch trajectories that randomly drift by horizontal and vertical amounts within

the 2D interval $[-R, R]^2$ from frame to frame.

Using the space-time displacement vector notation we introduced earlier, we denote vectors for the motion, non-displaced, and random trajectories by $d_M = (x_M(t), y_M(t), t)$, $d_{ND} = (0, 0, t)$, and $d_R = (x_R(t), y_R(t), t)$, respectively. Here, $(x_M(t), y_M(t))$ is a spatial displacement vector between the initial spatial coordinate $(i, j)$ at time instant $t_0$, and the spatial coordinate arrived at time instant $t_0 + t$ by tracing along the ground-truth motion trajectory. The $(x_R(t), y_R(t))$ is a spatial displacement vector between the initial spatial coordinate $(i, j)$ at time instant $t_0$ and the spatial coordinate arrived at time instant $t_0 + t$ by tracing along the random trajectory constructed from uniform randomly generated vectors in the 2D interval $[-R, R]^2$ at each progression of frames. Note that we assume the existence of a video for which there are known or estimated ground-truth motion vectors, which are used to define the motion trajectories. We used the freely available HD1K optical flow data set [28], which provides ground truth optical flow vectors on short (~1 sec) $2560 \times 1080$ videos. Figure 2 shows an example of patch tracing of three different space-time displacement trajectories. Frame-difference volumes are formed by collecting the frame-differenced patches (displaced or otherwise) along each trajectory.

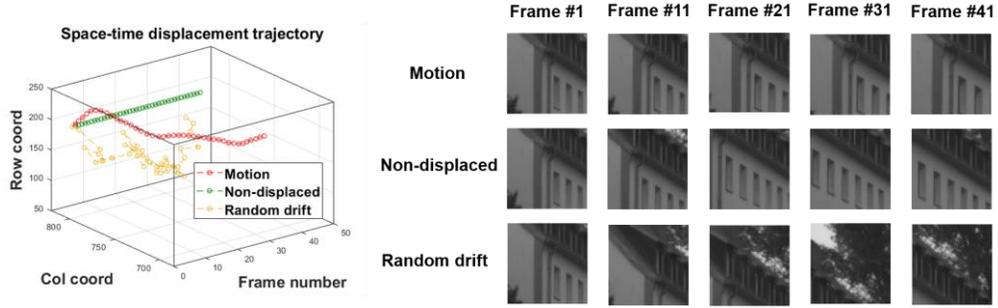

Fig. 2. Patches along various space-time displacement trajectories.

*3.2 Spatial/ Temporal/ Spatio-temporal divisive normalization*

Given three-dimensional space-time differenced patch volumes computed along the three kinds of trajectories, the next step is to apply spatial, temporal and spatio-temporal divisive normalization. The form of coefficients that we will compute on space-time displaced frame differences is given by

$$\hat{I}_{fd}(i, j, k) = \frac{I_{fd}(i, j, k)}{\sigma(i, j, k) + C}, \tag{5}$$

where $I_{fd}$ refers to an input volume of space-time displaced frame-differences, $i \in [1, w]$, $j \in [1, h]$, and $k \in [1, t]$, where $w$, $h$, and $t$ are the input volume dimensions, $\mu$ and $\sigma$ are locally gaussian-weighted space-time mean and contrast volumes, and where $C = 0.5$ is a stabilizing (saturation) constant. Note that (5) differs from (1), in that it is computed on a three-dimensional volume. Also, the local average is not subtracted from the numerator, since our input volume is space-time displaced frame-differences, which are band-pass coefficients that are already zero-centered.

We applied one of three types of divisive normalization to the input volumes. Figure 3 illustrates the local windows over which data is collected to compute the divisor when conducting temporal, spatial, and spatio-temporal divisive normalization.

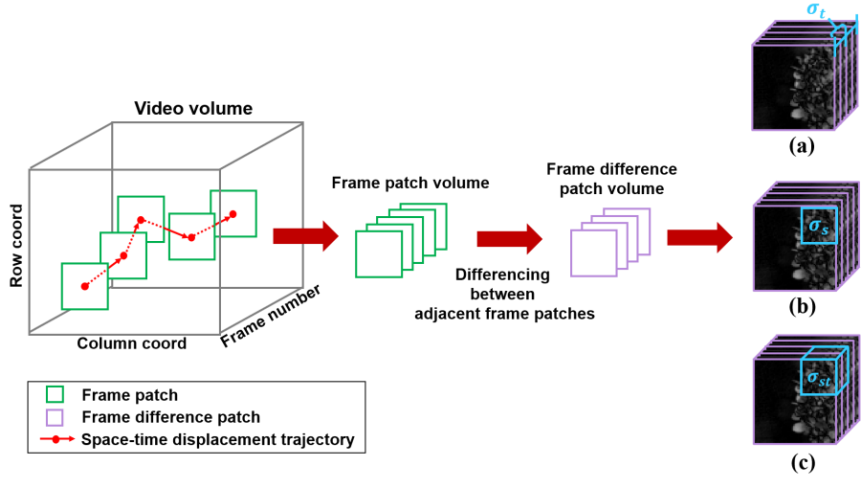

Fig. 3. Local windows over which contrast divisors are computed to conduct (a) temporal, (b) spatial, and (c) spatio-temporal divisive normalization.

Temporal divisive normalization (TDN) uses RMS contrast computed only along the 1-D temporal direction:

$$\sigma_t(i,j,k) = \sqrt{\sum_{n=-N}^{N} \omega_n [I(i,j,k+n) - \mu_t(i,j,k)]^2}. \qquad (6)$$

Spatial divisive normalization (SDN) uses contrast computed on 2-D spatial patches:

$$\sigma_s(i,j,k) = \sqrt{\sum_{l=-L}^{L} \sum_{m=-M}^{M} \omega_{lm} [I(i+l,j+m,k) - \mu_s(i,j,k)]^2}. \qquad (7)$$

Finally, spatio-temporal divisive normalization (STDN) uses contrast computed on space-time cuboids:

$$\sigma_{st}(i,j,k) = \sqrt{\sum_{l=-L}^{L} \sum_{m=-M}^{M} \sum_{n=-N}^{N} \omega_{lmn} [I(i+l,j+m,k+n) - \mu_{st}(i,j,k)]^2}. \qquad (8)$$

The window functions $\omega_n$, $\omega_{lm}$, and $\omega_{lmn}$ are 1D, 2D and 3D symmetric Gaussians sampled out to three standard deviations ($L = 5$, $M = 5$, $N = 10$) and rescaled to unit volume. The weighted mean luminance functions $\mu_t$, $\mu_s$, and $\mu_{st}$ computed over spatial, temporal and spatio-temporal intervals are:

$$\mu_t(i,j,k) = \sum_{n=-N}^{N} \omega_n I(i,j,k+n), \qquad (9)$$

$$\mu_s(i,j,k) = \sum_{l=-L}^{L} \sum_{m=-M}^{M} \omega_{lm} I(i+l,j+m,k), \qquad (10)$$

and

$$\mu_{st}(i,j,k) = \sum_{l=-L}^{L} \sum_{m=-M}^{M} \sum_{n=-N}^{N} \omega_{lmn} I(i+l,j+m,k+n). \qquad (11)$$

*3.3 Statistical regularity of displaced space-time frame differences*

Using the video data in the aforementioned HD1K database, we extracted a variety of frame-difference volumes containing $100 \times 100 \times 40$ space-time samples. These were captured along motion, non-displaced, and random trajectories. Then, we applied divisive normalization (5) using RMS contrast for TDN, SDN, or STDN ((6)-(8)), followed by normalization to unit sample variance (which is not guaranteed by the directional normalization), on each space-time volume.

Our canonical model assumes that the bandpass, normalized coefficients of videographic data, whether of frames or frame differences, obey a gaussian distribution law [20]. However, as we show here, this kind of regularity does not necessarily hold for frame differences, unless they are displaced in the direction of motion. Thus, to allow for departures from gaussianity, we will more generally model frame-difference data along the various space-time trajectories as following a generalized gaussian distribution (GGD) law. The two parameter GGD has density function

$$f(x;\alpha,\sigma^2) = \frac{\alpha}{2\beta\Gamma\left(\frac{1}{\alpha}\right)} \exp\left(-\left(\frac{|x|}{\beta}\right)^\alpha\right), \qquad (12)$$

where $\beta = s\sqrt{\frac{\Gamma(1/\alpha)}{\Gamma(3/\alpha)}}$, $\Gamma(a) = \int_0^\infty t^{a-1}e^{-t}dt$ $(a > 0)$ is the gamma function, $\alpha$ is the shape parameter that dictates the peakiness and tail weight of the GGD, and $s^2$ is the variance [29]. When $\alpha = 2$, the density becomes the canonical gaussian.

Figure 4 shows the best-fitting (ML) distribution fits to the empirical densities of the quantities (5) using the sample contrast and mean luminance functions (6)-(8), and (9)-(11), respectively. As shown in Fig. 4(a), each frame-difference patch volume was constructed by collecting and differencing patches traced along motion, non-displaced, and random trajectories. The random trajectories were generated as sequences of vectors uniformly sampled from $[-20, 20]^2$ along each progression of frames. The average magnitudes of the frame difference values provided in the table under the plot in Fig. 4(a) represent the residuals obtained by differencing adjacent frames along the different types of trajectories. As might be expected, the residuals are smaller along the motion trajectories, while much larger residuals occur along the random trajectories because of the frequent mismatches between patches that are differenced. Fig. 4(b) shows the distributions of frame difference coefficients along each trajectory, where significant differences in the best GGD fits, which are overlaid, may be observed. Since these have not been contrast normalized, they are not expected to be gaussian. The frame differenced coefficients computed along the ground truth motion trajectory present highly peaked distribution, with increased concentration about zero and sparser non-zero coefficients. The coefficient distributions along the non-displaced and random trajectories, by contrast, are less peaky and more spread out.

The histograms of frame difference coefficients following TDN are shown in Fig 4(c). Notably, the coefficient histogram computed along the random trajectory is highly irregular, while the histogram computed along the non-displaced frame difference trajectory is also not very Gaussian. However, the histogram of frame differences displaced along the motion trajectory is strongly Gaussianized. Going forward, we will be interested in quantifying this kind of regularity (or lack thereof). One way to quantify how well the TDN coefficients agree with the canonical gaussian model is by computing the Kullback Leibler Divergence (KLD) between the coefficient distribution and the canonical gaussian distribution. The KLD is given by [30]:

$$D_{KL}(P \| Q) = \sum_i P(i) \log\left(\frac{P(i)}{Q(i)}\right), \qquad (13)$$

where we take $P(i)$ and $Q(i)$ to be the empirical probability densities of the TDN coefficients and the canonical gaussian, respectively. For the data plotted in Fig. 4(c), we obtained KLD values of 0.0007, 0.0082, and 0.0400 for the motion, non-displaced, and random trajectories, respectively, which supports visual observation of the distributions: the TDN transformed coefficients are the most regular, and adhered best to the gaussian model, along space-time trajectories in the motion direction.

Figure 4(d) shows the distributions of SDN coefficients. Application of SDN produces very stable distributions, albeit with varying degrees of spread. The KLD between the histograms of SDN coefficients in Fig. 4(d) and the gaussian model were 0.0078, 0.0121, and 0.1443 for the motion, non-displaced, and random trajectories, respectively. The SDN coefficients again tended significantly more towards Gaussianity along the motion trajectory. However, the differences were reduced, primarily because SDN strongly reduces spatial content redundancies, which are present in regardless of the trajectory along which differences are taken.

The distributions of STDN coefficients are depicted in Fig. 4(e). Here we again observe model departures, especially along the random trajectory. The STDN distributions adhered much more closely to the gaussian model along the motion trajectory, where the KLD of the histograms of the STDN coefficients shown in Fig. 4(e) against the gaussian model were 0.0005, 0.0063, and 0.0251 for the motion, non-displaced, and random trajectories, respectively. Generally, the KLD values were reduced as compared to the TDN distributions owing to the strong influence of (partly) spatial normalization.

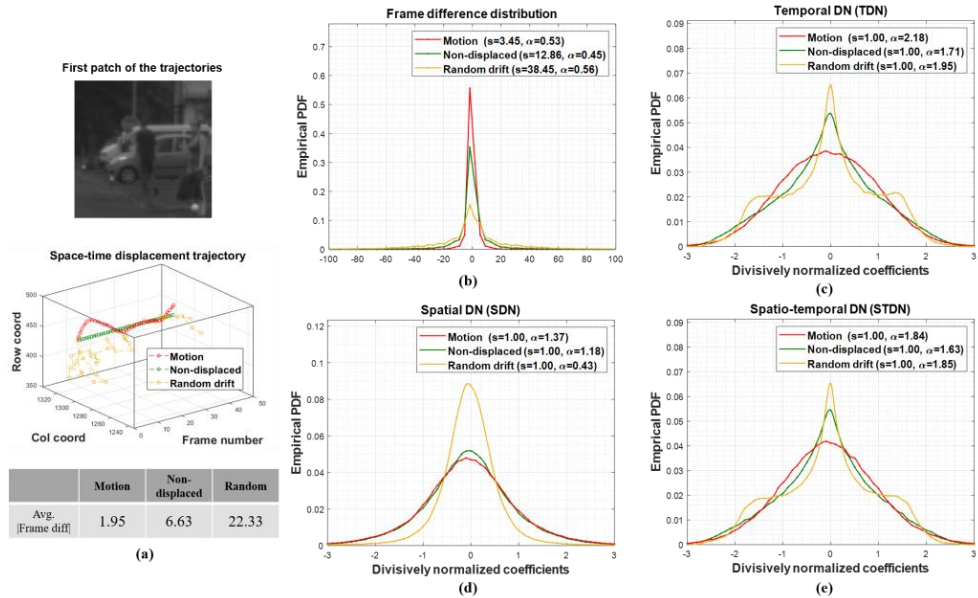

Fig. 4. Distributions of frame-differenced video coefficients before and after divisive normalization along different trajectories. (a) Plots of different space-time trajectories (motion, non-displaced, and random), and the average absolute frame difference values along each path. (b) Empirical distributions of frame difference coefficients without any normalization. (c) Distributions of TDN coefficients. (d) Distributions of SDN coefficients. (e) Distributions of STDN coefficients.

Figure 5 shows another example on a patch containing a textured region. Similar conclusions may be reached when interpreting Figs. 5(c)-5(e), where the coefficients along the motion trajectory tended more strongly toward gaussianity. The KLD values of the histogram of the TDN coefficients (Fig. 5(c)) against gaussianity were 0.0040, 0.0247, and 0.0160 (motion, non-displaced, and random trajectories). This again illustrates the strong divergence in the behavior of normalized coefficients along the motion trajectory as compared to the non-displaced and random trajectories. When using SDN (Fig. 5(d)), the KLD values between the computed coefficient histograms and the gaussian model were 0.0003, 0.0046, and 0.0228. Relatively small values of the KLD were obtained because of the strong spatial decorrelation imparted by SDN. However, this does not reveal the discrimination of statistical regularity along the space-time direction as clearly. Finally, when using STDN (Fig. 5(e)) the obtained KLD values between the computed coefficient histograms and the gaussian model were 0.0015, 0.0097, and 0.0145, respectively.

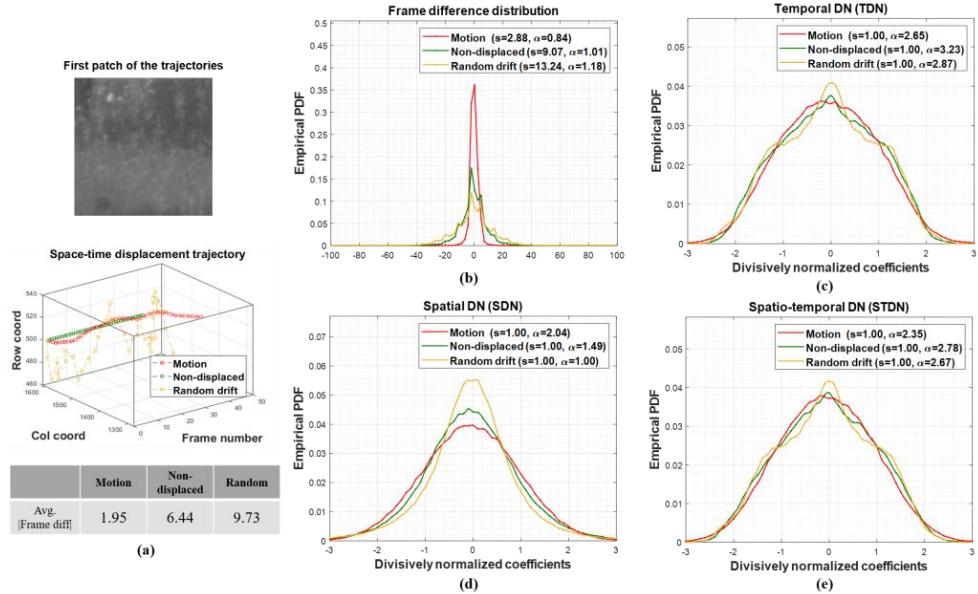

Fig. 5. Distributions of frame-differenced video coefficients before and after divisive normalization along different trajectories. (a) Plots of different space-time trajectories (motion, non-displaced, and random), and the average absolute frame difference values along each path. (b) Empirical distributions of frame difference coefficients without any normalization. (c) Distributions of TDN coefficients. (d) Distributions of SDN coefficients. (e) Distributions of STDN coefficients.

An intermediate and important conclusion, which we have found agreement with in all the videos we have studied, is that there are strongly, inherent statistical regularities in frame difference signals that are aligned along the motion direction. Applying data normalization along other trajectories produces much less regular, less Gaussian behavior. Overall, it is apparent that spatial normalization (SDN) increases the tendency towards Gaussianity regardless of the trajectory type, while temporal normalization (TDN) reveals that Gaussianity is maximized along the motion trajectory, but is lost along other trajectories, unless the amount of motion is small. Space-time normalization (STDN) captures the differences in regularity (adherence to gaussianity) amongst the trajectories quite well, while producing the most highly gaussianized coefficients. These differences in statistical regularity are certainly available to, and measurable by, neurons in the visual system, albeit likely in a very different way. Indeed, we may speculate that the responses of neuronal populations to the degree of regularity of bandpass space-time statistics may guide or influence motion computations further along the visual pathway, such as motion flow computation and tracking.

From the application perspective, these results and those to follow help explain the "motion compensation" aspect of digital video compression, whereby efficiency is greatly enhanced by coding entropy-reduced (and more regular) residuals of bandpass discrete cosine transform coefficients along "motion compensated" motion trajectories. This also suggests that these regularities could be used to predict motion fields (as we shall shortly see), but without conducting millions of metric-based matching searches. These results also relate to optical flow models, where redundancy across temporal sequences is expressed in terms of both local luminance constancy along the motion trajectories [31-34] and gradient (contrast) constancy [35].

We hypothesize that by computing the histograms of space-time displaced frame data, it is possible to differentiate motion trajectories from non-motion trajectories, an idea that we will explore in detail in Section 4. This is borne out in Fig. 6, which plots the histograms of displaced frame difference coefficients for all three types of trajectories, when subjected to TDN, SDN, and STDN and normalized to unit sample variance. Overlaid are plots of the model $\mathcal{N}(0,1)$ densities.

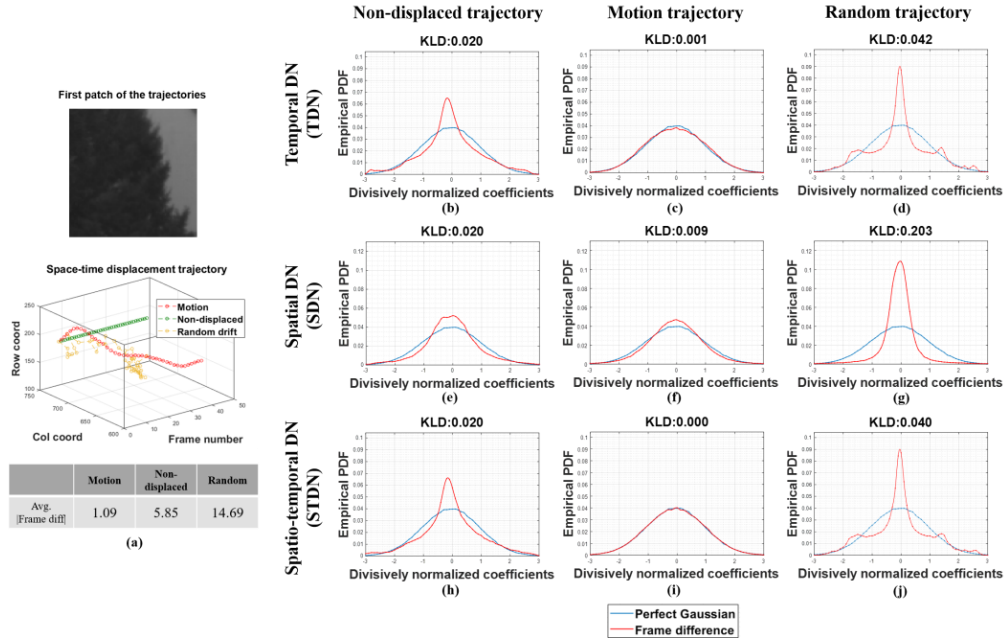

Fig. 6. Comparison of distributions between perfect Gaussian and frame-difference displaced in various trajectories (motion, non-displaced and random) subject to various divisive normalization (temporal, spatial, and spatio-temporal).

The preceding results are strongly indicative that, among the many possible space-time displacement trajectories, along which localized frame differencing may be computed, ones aligned in the direction of motion best capture the temporal statistical regularities inherent in motion pictures. Of course, this is not entirely surprising, given the efficacy of video coding systems that exploit temporally displaced redundancies to create low-entropy, sparse, highly compressible representations. More surprising, however, is the degree to which the phenomena holds. In fact, as we show next, simply finding space-time paths that are statistically regular is an effective way to estimate motion flow.

## 4. Motion flow estimation from space-time statistical regularity

To further test the proposition that space-time displacement trajectories that preserve statistical regularities coincide with motion, we developed a way to estimate the motions of patches, using easily-computed space-time regularity maps. We show the estimates are accurate, using a well-known benchmark optical flow database. While this was intended as an exercise to test our hypothesis regarding the regularity of motion trajectories, we were surprised to find that the resulting "motion prediction algorithm" is competitive with these classical models.

### 4.1 Motion estimation method

Our method of motion estimation utilizes frame difference coefficients calculated along various space-time trajectories. Motion paths are computed by seeking the space-time displacement direction of "least statistical resistance," i.e., the most regular path, using the KLD as the fundamental tool.

Figure 7 diagrams the way we construct space-time regularity maps. First, partition each video frame into patches of size $N \times N$. We then generated space-time displaced patch frame differences using Equation 4, and computed the histogram of each (scaled to sum to unity). We constrained the displacement vector $d = (x, y, t)$ to the range $[-R, R]^2$, defined relative to the patch size according to $R = \lfloor N/6 \rfloor - (\lfloor N/6 \rfloor \mod 2)$. This reduced complexity, and is supported by studies that have shown that inter-frame velocity is generally small [36-38]. Second, we fixed the temporal displacement to be $t = 1$. While statistical regularities are no doubt preserved over longer temporal separations, which are also very likely useful, we are able to show good performance using only adjacent frame differences. Since we compute space-time regularity maps only using adjacent frames, our model motion algorithm only deploys the normalization using SDN. However, as maybe seen from Fig. 6, the statistical regularity of the frame differences is preserved along the motion direction when using SDN. The space-time motion trajectory of a patch is thus estimated by collecting consecutive 'frame-wise' regularity-preserving motion vectors.

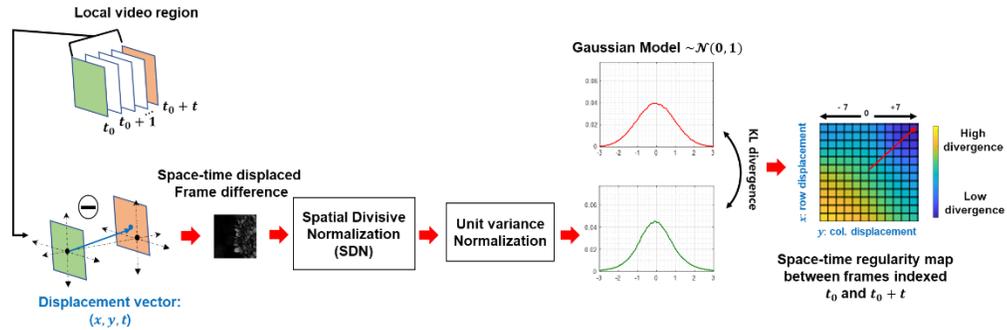

Fig. 7. How we constructed space-time regularity maps, which we then used to find space-time paths that best preserve the inherent statistical regularities of natural videos.

Each space-time regularity map measures the proximity between the reference Gaussian distribution and the space-time displaced distribution using the KLD defined in Equation 13. Since the displacements vary within $[-R, R]^2$, the space-time regularity maps, which are of dimensions $(2R+1) \times (2R+1)$, contain the KLD value for each displacement coordinate, as shown in Figure 7. The motion direction for the selected patch is estimated as the average of those displacement vectors yielding KLD values less than 95% (lowest 5th percentile) of the values in the space-time regularity map. This vector is part of an optimal space-time path that best preserves the space-time bandpass regularity in the video.

Fig. 8 shows examples of this regularity-maximizing motion estimation process. Figs. 8(a), (d), and (g) depict local regions in video frames from the Middlebury optical flow benchmarking database, along with (heavily downsampled, for visualization) respective average ground truth motion vectors. Figs. 8(b), (e), and (h) show the space-time regularity map computed using our model, and optimal space-time paths computed from them (the displacement indicated by the deepest blue). In all cases, the optimal space-time path appears to nicely approximate the motion of patches. Of course, a better evaluation can be made my quantitative comparison with ground-truth motion vectors, an analysis that we conduct comprehensively in the next section. Note that in Figs. 8(g) and (h), the regularity map and motion estimates were computed using a temporal frame separation $t = 2$, where the motion vector was (-3,2). Comparing pairs of frames separated by $t$ unit frame intervals, the resulting motion vectors may be expressed as three-dimensional vectors $(-3 \times t, 2 \times t, t)$. In this particular example, the resulting ground truth MV may be denoted (-6,4,2). As shown in Fig. 8(h), the optimal space-time path is (-6,6,2), which approximates the apparent motion direction. Figs. 8(c), (f), and (i) portray the distribution of SDN coefficients on frame differences without any spatial displacements (FD non-displaced), frame differences displaced in the motion direction (FD motion), and frame-differences displaced in the opposite direction as the motion direction (FD opposite), all of which are compared against the ideal Gaussian model.

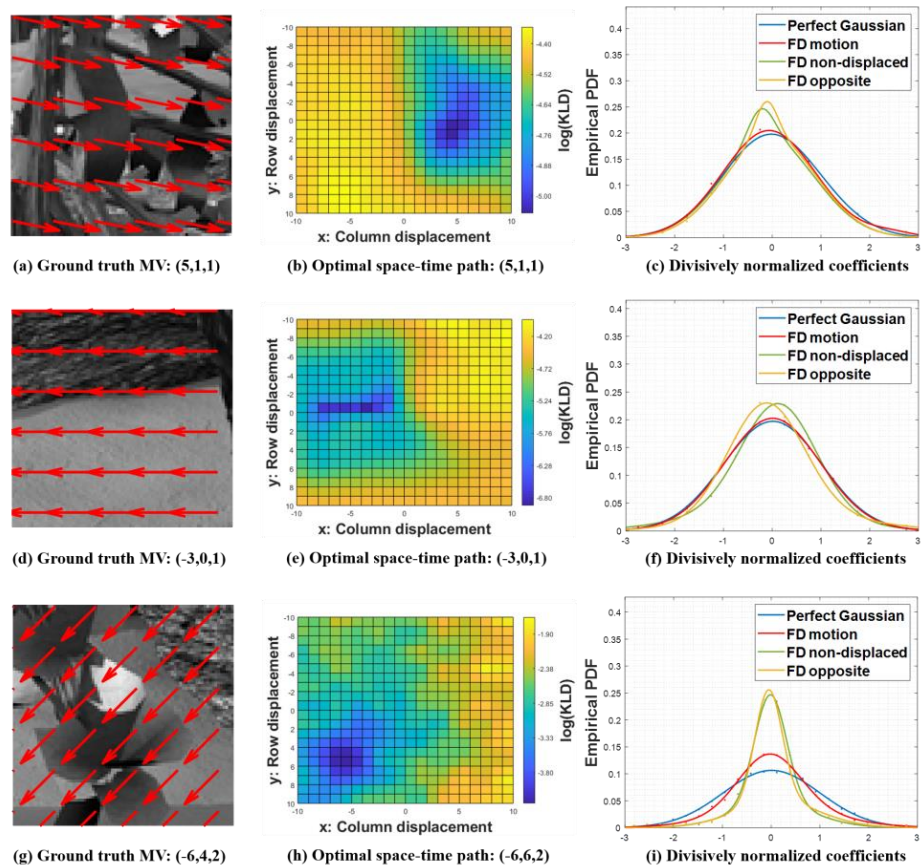

Fig. 8. (a), (d), (g) Video patches from the Middlebury optical flow database sequences, with the average ground truth motion vectors of each patch indicated by red arrows. (b), (e), (h) Space-time regularity maps, with the optimal regular space-time path indicated by the darkest color with value given by the label. (c), (f), (i) Distributions of perfect gaussian model, SDN coefficients for non-displaced frame-differences (FD non-displaced), frame-differences displaced in the motion direction (FD motion), and frame-differences displaced in the direction opposite of motion (FD opposite)

As expected, all of the computed distributions exhibit some degree of statistical regularity (because of the spatial normalization), but the distribution of frame differences displaced in the motion direction was most similar to the model distribution. As shown in Fig.8(c), the KLD values of the space-time displaced frame difference distributions in the motion, non-displaced, and opposite of motion directions were 0.0041, 0.0095, and 0.0122, respectively. In Fig. 8(f), the KLD values of the space-time displaced distributions in the motion, non-displaced, and opposite of motion directions with respect to the model gaussian distribution were 0.0011, 0.0096, and 0.0108, respectively. Finally, in Fig. 8(i), the KLD values of the space-time displaced distributions in the motion, non-displaced, and opposite of motion directions with respect to the model gaussian distribution were 0.0204, 0.1282, and 0.1420, respectively. These examples show that there exist displacement paths that best preserve the space-time statistical regularity of frame differences, and that these "regularity trajectories" are highly correlated with the direction of image motion.

We also present a coarse-to-fine method of estimating the motion trajectory of a patch using TDN or STDN coefficients. Figure 9 diagrams the procedure for evaluating the statistical regularity of the TDN or STDN coefficients along a certain space-time displacement trajectory. To estimate the motion trajectory of a $100 \times 100$ patch at time instant $t_0$, we first fixed a set of 25 "initial" displacement vectors $(x, y)$ from the search range $[-24, 24]^2$ all arriving at time instant $t = t_0 + 10$ (frames). These were selected so that the endpoints of the initial vectors in the $(t_0 + 10)$ TH frame (relative to coordinate $(x, y)$ in frame indexed to) were equally spaced by 12 pixels in the cardinal directions. Then, for each initial vector, we constructed a frame difference patch volume by collecting frame difference patches along the straight-line trajectory in $(x, y, t)$ dictated by the initial vector. Hence, we assumed that over very short intervals (10 frames), motion trajectories can be approximated as straight lines. Each resulting frame difference patch volume of dimension $100 \times 100 \times 10$ was then subjected to TDN (or STDN), using local window sizes of $N = 5$, and $L = 5$, $M = 5$, and $N = 5$ for TDN and STDN, respectively. The TDN or STDN coefficients were then subjected to unit variance normalization, and the resulting distributions of coefficients compared against the ideal Gaussian model. After finding the trajectory yielding the lowest KLD value, then, proceeding in coarse-to-fine manner, a set of 9 (instead of 25) second-stage vectors were selected in the vicinity of the selected vector, yielding a new selected vector, and the decision process repeated. This second set of steps was repeated twice more, achieving finer accuracy with each step, ultimately yielding an efficient "4-step" solution that avoided exhaustive search.

The motion estimation process then proceeded to the next frame, until the process was stopped at the end of the video.

The four-step search process is similar to the three-step search [37] method used in traditional motion estimation models. The grid distances were decreased to 12, 6, 3, and 1 over the four steps.

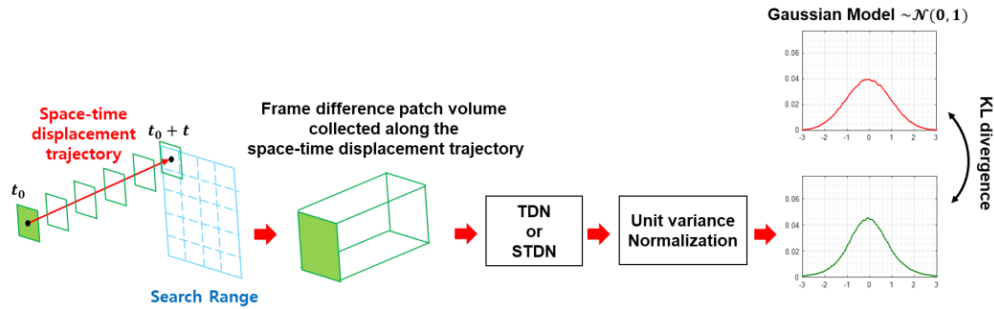

Fig. 9. Procedures for measuring the statistical regularity of TDN or STDN coefficients collected along a given space-time displacement trajectory.

### 4.2 Motion estimation accuracy

Given these promising results, we were motivated to assess the performance of our simple regularity-based motion prediction model on a larger set of ground-truth motion flow data. While our goal has been to explore an interesting statistical regularity property of naturalistic motion pictures, and not to create a competitive "computer vision" algorithm, we also compared our model with some well-known, canonical optical flow algorithms [31,34].

### 4.2.1 Comparison methods

We studied the performance of our method on a standard optical flow benchmark, and also compared its performance with two other well-known motion estimation methods. One is the Horn-Schunck method, which is a classical iterative optical flow algorithm that is based on the luminance constancy assumption and a flow smoothness constraint [31]. Horn-Schunck yields a dense map of flow vectors. The other method is the Black-Anandan algorithm, which significantly improves upon Horn-Schunck by modifying the luminance constancy loss function by introducing concepts of robust statistics [34]. Both Horn-Schunck and Black-Anandan have been regarded as top performers in past computer vision studies. Of course, algorithm criteria in that field is very different than perceptual relevance, but in any case, will serve as good comparators for evaluating our perceptually motivated motion estimator.

### 4.2.2 Evaluation metrics

We used the de facto standard angular error (AE) and endpoint error (EE) to measure the accuracy of the predicted motion vectors. The AE measures the angular distance between the estimated motion vector $(u_{est}, v_{est}, 1.0)$ and the ground-truth motion vector $(u_{GT}, v_{GT}, 1.0)$, expressed in terms of the normalized dot products the predicted and ground truth motion vectors [32,33]. The AE is

$$AE = \cos^{-1}\left(\frac{u_{est}u_{GT} + v_{est}v_{GT} + 1}{\sqrt{(u_{est}^2 + v_{est}^2 + 1)}\sqrt{(u_{GT}^2 + v_{GT}^2 + 1)}}\right). \quad (14)$$

The EE metric provides complementary performance, and is formulated as [39]:

$$EE = \sqrt{(u_{est} - u_{GT})^2 + (v_{est} - v_{GT})^2}, \quad (15)$$

which is the Euclidian distance between the endpoints of the estimated and ground truth motion vectors.

We also report the actual computation time elapsed by each of the motion estimation methods, while varying parameters to obtain a sense of the potential practicality of our model. The computation times are provided in unit of seconds expended to estimate the motion vectors on a single video frame. All of the methods were implemented in MATLAB, using the implementations of Sun et al. [40] with parameters adjusted to closely follow the original algorithms of Horn & Schunck [31] and Black & Anandan [34]. All of the methods were run on a system with an Intel i7-8700 CPU@3.2GHz and 32GB RAM. Practical implementations could be tremendously accelerated, of course using efficient programming and hardware.

### 4.2.3 Performance results

Our main goal was to determine whether our perceptually motivated space-time statistical regularity model can produce motion estimates that are close to true motion.

We first present our results on the motion estimation method using SDN coefficients. Since our proposed regularity-based motion estimator uses patches, there will naturally exist trade-offs between patch size and prediction accuracy. Hence, we varied the patch sizes from 51×51 to 101×101 in increments of 10. Horn-Schunck was iterated 6, 12 and 18 times. Black-Anandan was set at pyramid level 2 and iterated 2, 3, and 4 times. Table 1 and Figure 10 show the angular error performance results for all models, while Table 2 and Figure 11 show the endpoint error performance results on the Middlebury optical flow benchmarking database [41]. Overall, our proposed method attained very comparable motion estimation performance against the other methods, while requiring less computation.

There was an observable effect on performance accuracy of patch size, although it varied with video content. As shown in Figure 12, very small patch sizes. which may not encompass meaningful structures or textures, may have misled the proposed predictor in places. Generally, increasing the patch size improved prediction performance. However, in some instances, increasing the patch size led to small decreases in performance.

The 'Rubberwhale' sequence is a good study of the limits of the model, on which somewhat worse prediction performance was obtained as compared to the other motion estimation methods. This sequence contains many local areas containing opposing motion directions, hence processing large patches may reduce regularities present over smaller scales.

Very good performance was obtained on the 'Urban3' sequence, better than the other compared methods. This sequence contains significant structural information, such as buildings and windows. The proposed method was able to exploit this highly regular content. The error maps in Figures 10 and 11 show that the other methods performed relatively poorly on the structured regions, whereas the proposed regularity-preserving method performed well.

Table 3 shows the motion trajectory prediction performance results on the HD1K optical flow benchmarking database. The HD1K database provides uncertainty labels on videos containing challenging conditioning, such as low-light, rain, and distortion, that affect the accuracy of ground truth motion values. We computed the results of all algorithms only on patch volumes without uncertain regions. The methods compared here are the proposed motion trajectory estimation using TDN and STDN coefficients, the Black-Anandan model set at pyramid level 2 iterated 3 times, and Horn-Schunck iterated 12 times. We see from the overall EE results that both of our trajectory estimation methods (based on TDN and STDN coefficients) delivered competitive performances against the other methods.

Table 1. Angular Error (AE) and Computation Time of Compared Motion Estimation Models

| Methods | Parameters | | Contents | | | | | | | | | | | |
|---|---|---|---|---|---|---|---|---|---|---|---|---|---|---|
| | | | Grove2 | | Grove3 | | Hydrangea | | Rubberwhale | | Urban2 | | Urban3 | |
| | Patch size | Displacement range | AE | Time [sec] | AE | Time [sec] | AE | Time [sec] | AE | Time [sec] | AE | Time [sec] | AE | Time [sec] |
| Proposed (SDN) | 51x51 | [-8, 8] | 36.66 | 7.12 | 27.35 | 7.49 | 27.78 | 5.94 | 43.09 | 5.67 | 72.33 | 8.78 | 33.55 | 7.32 |
| | 61x61 | [-10, 10] | 30.21 | 7.53 | 30.35 | 7.61 | 27.87 | 6.43 | 40.35 | 5.97 | 71.23 | 7.61 | 29.66 | 7.50 |
| | 71x71 | [-10, 10] | 25.62 | 6.36 | 30.91 | 6.17 | 17.41 | 4.60 | 37.60 | 4.64 | 65.88 | 6.05 | 31.11 | 6.53 |
| | 81x81 | [-12, 12] | 23.71 | 5.90 | 25.44 | 5.79 | 22.10 | 4.77 | 33.73 | 4.84 | 67.44 | 6.08 | 27.52 | 5.77 |
| | 91x91 | [-14, 14] | 17.63 | 7.96 | 27.26 | 7.97 | 23.04 | 5.56 | 43.01 | 5.46 | 59.07 | 8.19 | 21.48 | 8.12 |
| | 101x101 | [-16, 16] | 18.84 | 7.76 | 32.84 | 7.21 | 22.58 | 4.80 | 36.59 | 4.54 | 70.34 | 7.24 | 26.98 | 7.72 |
| Horn-Schunck | | *iter=6* | 42.01 | 8.60 | 37.52 | 8.59 | 19.45 | 6.34 | 11.25 | 6.31 | 54.19 | 8.59 | 43.02 | 8.54 |
| | | *iter=12* | 33.62 | 17.14 | 28.21 | 17.20 | 8.64 | 12.57 | 10.95 | 12.57 | 50.14 | 17.09 | 33.77 | 16.93 |
| | | *iter=18* | 31.10 | 25.61 | 23.52 | 25.77 | 8.11 | 18.88 | 10.85 | 18.80 | 47.93 | 25.77 | 29.29 | 25.31 |
| Black-Anandan | | *pyr=2, iter=2* | 23.90 | 19.20 | 19.56 | 19.22 | 8.74 | 13.90 | 10.32 | 13.88 | 48.56 | 19.26 | 27.63 | 19.17 |
| | | *pyr=2, iter=3* | 21.25 | 28.61 | 16.71 | 29.60 | 8.32 | 21.13 | 9.97 | 20.39 | 47.15 | 28.50 | 23.41 | 28.87 |
| | | *pyr=2, iter=4* | 18.51 | 38.35 | 15.11 | 38.15 | 8.07 | 27.45 | 9.73 | 27.36 | 45.71 | 37.56 | 20.99 | 37.49 |

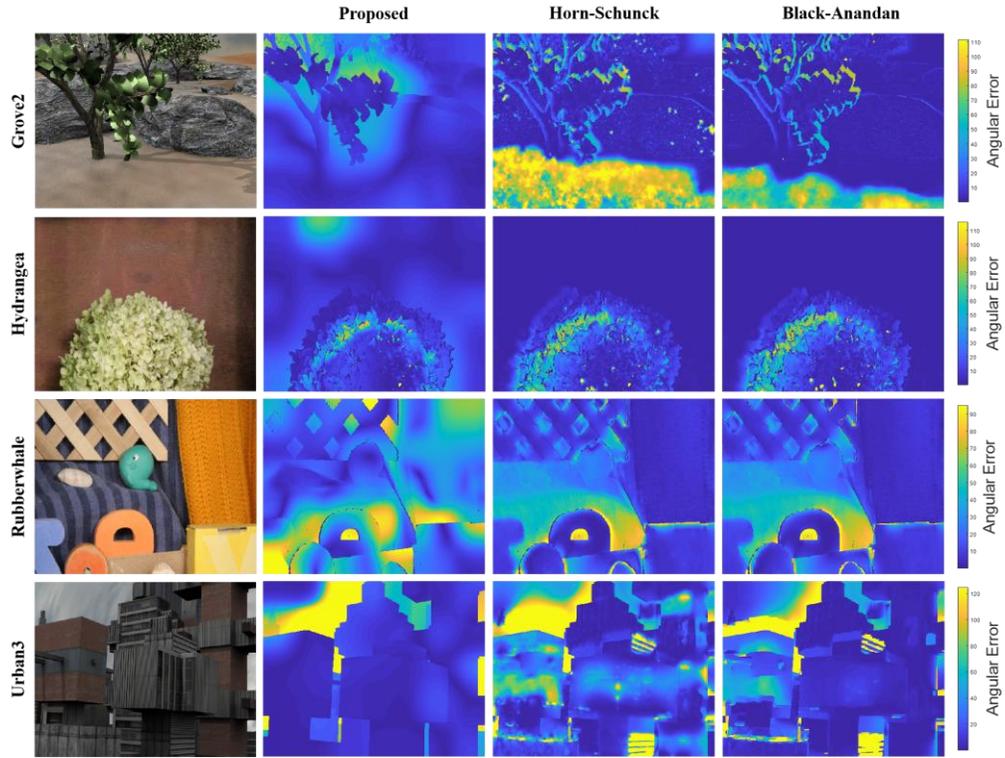

Fig. 10. Angular error maps for proposed, Horn-Schunck, and Black-Anandan methods.

**Table 2. Endpoint Error (EE) and Computation Time of Compared Motion Estimation Models**

| Methods | Parameters | | Contents | | | | | | | | | | |
|---|---|---|---|---|---|---|---|---|---|---|---|---|---|
| | | | Grove2 | | Grove3 | | Hydrangea | | Rubberwhale | | Urban2 | | Urban3 | |
| | Patch size | Displacement range | EE | Time [sec] | EE | Time [sec] | EE | Time [sec] | EE | Time [sec] | EE | Time [sec] | EE | Time [sec] |
| Proposed (SDN) | 51x51 | [-8, 8] | 1.96 | 7.12 | 2.33 | 7.49 | 2.21 | 5.94 | 1.43 | 5.67 | 8.98 | 8.78 | 4.47 | 7.32 |
| | 61x61 | [-10, 10] | 1.80 | 7.53 | 2.61 | 7.61 | 2.12 | 6.43 | 1.36 | 5.97 | 8.91 | 7.61 | 3.65 | 7.50 |
| | 71x71 | [-10, 10] | 1.60 | 6.36 | 2.87 | 6.17 | 1.64 | 4.60 | 1.16 | 4.64 | 8.49 | 6.05 | 3.77 | 6.53 |
| | 81x81 | [-12, 12] | 1.57 | 5.90 | 2.82 | 5.79 | 2.02 | 4.77 | 1.23 | 4.84 | 7.72 | 6.08 | 3.49 | 5.77 |
| | 91x91 | [-14, 14] | 1.40 | 7.96 | 2.95 | 7.97 | 2.06 | 5.56 | 1.58 | 5.46 | 7.71 | 8.19 | 2.87 | 8.12 |
| | 101x101 | [-16, 16] | 1.42 | 7.76 | 3.08 | 7.21 | 2.22 | 4.80 | 1.40 | 4.54 | 8.05 | 7.24 | 3.27 | 7.72 |
| Horn-Schunck | iter=6 | | 2.00 | 8.60 | 2.95 | 8.59 | 1.84 | 6.34 | 0.38 | 6.31 | 7.97 | 8.59 | 5.91 | 8.54 |
| | iter=12 | | 1.54 | 17.14 | 2.49 | 17.20 | 0.71 | 12.57 | 0.36 | 12.57 | 7.81 | 17.09 | 5.22 | 16.93 |
| | iter=18 | | 1.42 | 25.61 | 2.23 | 25.77 | 0.68 | 18.88 | 0.36 | 18.80 | 7.67 | 25.77 | 4.77 | 25.31 |
| Black-Anandan | pyr=2, iter=2 | | 1.13 | 19.20 | 2.01 | 19.22 | 0.70 | 13.90 | 0.34 | 13.88 | 7.82 | 19.26 | 4.84 | 19.17 |
| | pyr=2, iter=3 | | 1.02 | 28.61 | 1.73 | 29.60 | 0.68 | 21.13 | 0.33 | 20.39 | 7.75 | 28.50 | 4.27 | 28.87 |
| | pyr=2, iter=4 | | 0.91 | 38.35 | 1.56 | 38.15 | 0.68 | 27.45 | 0.32 | 27.36 | 7.61 | 37.56 | 3.84 | 37.49 |

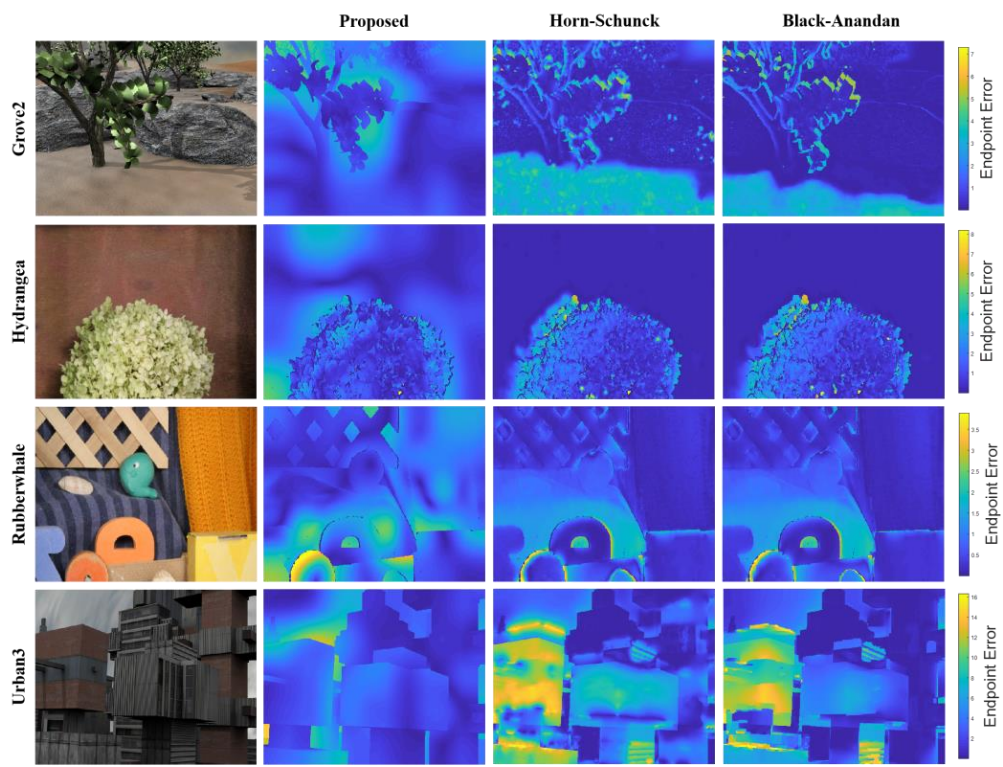

Fig. 11. Optical flow endpoint error maps of proposed, Horn-Schunck, and Black-Anandan method on four Middlebury videos.

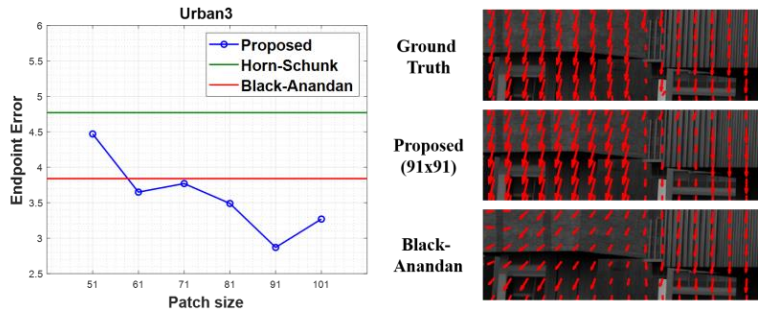

Fig. 12. Endpoint error plotted against varying patch sizes (left) and motion vector quiver plot (right)

Figure 13 shows examples of the motion trajectory estimation results on patches containing significant structural information, such as automobiles and tree branches. Again, our proposed method delivered performance comparable to the Black-Anandan method.

Table 3. Angular Error (AE) and Endpoint Error (EE) of the Estimated Motion Trajectories on the HD1K optical flow benchmark database

| Methods | Contents | | | | | | | | | | Overall | |
| --- | --- | --- | --- | --- | --- | --- | --- | --- | --- | --- | --- | --- |
| | 000000 | | 000011 | | 000016 | | 000026 | | 000030 | | | |
| | AE | EE | AE | EE | AE | EE | AE | EE | AE | EE | AE | EE |
| Horn-Schunck | 12.29 | 9.29 | 11.90 | 11.73 | 14.22 | 12.57 | 14.80 | 11.73 | 24.19 | 13.93 | 14.45 | 11.65 |
| Black-Anandan | 12.51 | 9.34 | 11.89 | 10.64 | 14.31 | 11.23 | 14.90 | 11.15 | 23.10 | 13.18 | 14.42 | 10.93 |
| Proposed (TDN) | 25.36 | 13.09 | 37.73 | 15.37 | 32.91 | 19.20 | 34.11 | 19.15 | 43.26 | 15.80 | 34.51 | 16.79 |
| Proposed (STDN) | 28.50 | 14.19 | 36.32 | 15.26 | 32.55 | 21.05 | 33.59 | 18.79 | 38.46 | 15.77 | 33.92 | 16.99 |
| Non-displaced | 67.41 | 36.43 | 69.71 | 40.38 | 70.12 | 31.61 | 66.88 | 26.08 | 58.67 | 26.08 | 67.29 | 36.69 |

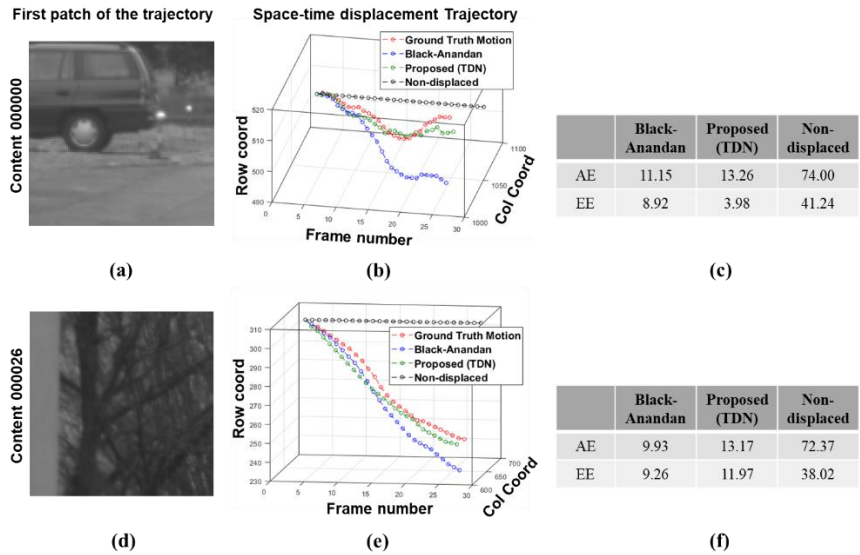

Fig. 13. Motion trajectory estimation results for proposed, Horn-Schunck, and Black-Anandan methods.

While our goal in here was not to create a new motion estimation model, but rather to explore the practicality of finding and exploiting statistical regularities in space-time visual data, the demonstrated motion estimator performs well, helping to establish our overall premise. It is simple and fast, does not require expensive matching computations, and is able to predict motion at a level comparable to classical and successful optical flow estimation methods.

## 5. Discussion and concluding remarks

We investigated the space-time statistics of natural motion pictures by observing and analyzing the probability distributions of displaced space-time frame difference along motion and non-motion trajectories. While we do not suggest that the new predictor is competitive with highly sophisticated computer vision models, which involve strong constraints, assumptions, search processes, or deep learning, it may help improve our understanding of how flow is processed in the brain. We believe that our findings may be relevant to motion processing in the early visual system, and how the brain has evolved in response to the statistics of a dynamic visual field.

Photographic and motion picture engineers may also be interested in these findings. The distributions of non-displaced frame difference have previously been used to drive highly successful perceptual video quality prediction engines [18-20], and of course, a foundation of video coding is by quantization of bandpass transformed frame difference coefficients to access their natural sparsity.

It is possible that our results on the statistical regularity of displaced space-time frame differences may further elevate the performances of these algorithms, or others designed for tracking, video stabilization, and other tasks. These results may also help us understand the role of micro-saccadic eye movements, which are recently implicated in the efficient coding of visual data near the point of gaze.

## Acknowledgements

This work was supported by the Institute for Information & Communications Technology Promotion (IITP) grant funded by the Korean government (MSIT) (No. 2017-0-00072, Development of Audio/Video Coding and Light Field Media Fundamental Technologies for Ultra Realistic Teramedia).

## Disclosures

The authors declare no conflicts of interest.